\documentclass[useAMS]{mn2e}
\usepackage{graphicx}
\usepackage{amsmath}

\title{KINETIC PROPERTIES OF FRACTAL STELLAR MEDIA}
\author[O. Chumak, A. Rastorguev]{
O.V.~Chumak$^{1}$\thanks{Deceased 2016 July, 13},
A.S.~Rastorguev$^{1,2}$;\\
$^{1}$ M.V.~Lomonosov Moscow State University, Sternberg
Astronomical Institute, 119992, 13,
Universitetskii prospect, Moscow, Russia\\
$^{2}$ M.V.~Lomonosov Moscow State University, Faculty of Physics,
119991, 1, Bld. 2, Leninskie Gory, Moscow, Russia;
alex.rastorguev@gmail.com}

\begin{document}

\date{Received 2016 February, 16; Accepted 2016 September, 30}

\pagerange{\pageref{firstpage}--\pageref{lastpage}} \pubyear{2016}

\maketitle

\label{firstpage}
\begin{abstract}
Kinetic processes in fractal stellar media are analysed in terms
of the approach developed in our earlier paper (Chumak \&
Rastorguev, 2015) involving a generalization of the nearest
neighbour and random force distributions to fractal media.
Diffusion is investigated in the approximation of scale-dependent
conditional density based on an analysis of the solutions of the
corresponding Langevin equations. It is shown that kinetic
parameters (time scales, coefficients of dynamic friction,
diffusion, etc.) for fractal stellar media can differ
significantly both qualitatively and quantitatively from the
corresponding parameters for a quasi-uniform random media with
limited fluctuations. The most important difference is that in the
fractal case kinetic parameters depend on spatial scale length and
fractal dimension of the medium studied. A generalized kinetic
equation for stellar media (fundamental equation of stellar
dynamics) is derived in the Fokker-Planck approximation with the
allowance for the fractal properties of the spatial stellar
density distribution. Also derived are its limit forms that can be
used to describe small departures of fractal gravitating medium
from equilibrium.
\end{abstract}


\section{Introduction}
The earliest notions about hierarchical, fractal structure of the
world of galaxies date back to Shapley (1934) and Carpenter (1938),
who showed that spatial number density in clusters of galaxies obeys
a power-law relation $ n(r)\sim r^{-\alpha} $ with the exponent of
about $\alpha \approx 1.5$, which is the same within the errors for
clusters of different dimensions. Later, de Vaucouleurs (1970)
concluded that the spatial distribution of galaxies obeys a
universal power-law relation, albeit with a different exponent of
$\alpha \approx 1.7$. Based on his results, Mandelbrot (1977) proposed
a hierarchical fractal model to describe the distribution of
galaxies. It became gradually clear that such extragalactic
structures as clusters and superclusters, walls, pancakes, etc., are
not base elements but rather parts of a general stochastic hierarchy
spanning a huge range of scales. In his fundamental studies Peebles
(1980, 1993) finally proved that density distribution in all
clusters of galaxies, irrespectively of their sizes, exhibits
properties that can be described in terms of two-point correlation
functions. Such functions have the form of power-law relations with
exponent $\alpha\approx1.7$, which is approximately the same for all
scales, directions, and distances.

Davis and Peebles (1983) introduced for power-law density
distribution the characteristic size $r_{0}$ defined by relation
$n(r_{0}) = 1$, which later came to be called the "correlation
length". This quantity is a matter of convention, because the actual
correlation length for power-law density distributions is infinite,
unlike the correlation length in molecular media or Debye radius in
plasma. Because of the power-law nature of the density distribution,
any definition of the size of the gravitating system is also only a
matter of convention, because all gravitating systems actually
correlate with each other and the nearest systems transit
smoothly into each other, featuring the overall fractal pattern of
the gravitating medium with a certain fractal dimension. However, in
practise the "correlation length" $r_{0}$ proposed by the above
authors proved to be useful for various numerical estimates.

Subsequent studies showed that this self-similar stochastic
hierarchy treated in terms of fractal concepts can describe quite
well not only the observed features of the world of galaxies
(Einasto, 1989; Klypin et al., 1989; Labini et al., 1996; Joyce et
al., 1999), but also those of gas-and-dust nebulae where star
clusters and associations form (Larson, 1981). The fractal model
of the distribution of galaxies was also convincingly confirmed by
direct numerical simulations (see Pietronero et al., 2002;
Pietronero \& Labini, 2005; Labini \& Pietronero, 2007 and
references therein). However, researchers have noted the lack of
spatial resolution in three-dimensional models of cosmological
gravitational clustering, which complicates the study of their
fractal properties down to smallest spatial scales. From the other
hand, it is possible to derive exact solutions for one-dimensional
cosmological clustering models. The 1D model was first formulated
by Rouet et al. (1990). More recent studies (see, for example,
Miller \& Rouet, 2010; Joyce \& Sicard, 2011; Benhaiem et al.,
2013) have shown that power spectrum of initial perturbations in
1D models leads to fractal and multifractal structures in a wide
range of scales. Detailed review of the problem can be found in
Shiozawa \& Miller (2016). One-dimensional cosmological models
seem to give correct qualitative understanding of many events in
the gravitational collapse of three-dimensional universe, but some
ambiguity still remains as to how to connect the properties of 1D
and 3D models (Joyce \& Sicard, 2011).

De Vega et al. (1998) demonstrated the fundamental differences
between the physics of gravitating and quasi-uniform media. The
above authors studied the thermodynamics of self-gravitating systems
using the theory of critical phenomena to show that gravity provides
a dynamical mechanism to produce finite multiscaled structures. This
{\it inherent} inhomogeneity of gravitating media, which has the
form of multiscaled fractal and stochastic hierarchy, represents a
fundamental difference between gravitating and quasi-uniform
systems.

A specific feature of such stochastic fractals is the self-
similarity of stochastic structures and the fact that the density
$n(r)$  of the distribution of structure elements does not tend to
any particular finite value. This actually means that there is no
such distance $r_{c}$ that density fluctuations $\Delta n(r)$  could
be neglected at $r>r_{c}$, i.e., $\Delta n(r)/n(r)$  does not vanish
at any $r$.

This is due to the fact that voids, which contain practically no
elements of the structure, are present at all observed scales, as is
immediately apparent, e.g., in the Sloan Digital survey data (SDSS).
This alternation property can be explained by the fact that the
statistics of fractals is not Gaussian. Fractal statistics
demonstrates other, non-Gaussian steady distributions (Mandelbrot,
1987). Because of these important differences, fractal statistics,
in particular, has to be taken into account in the analysis of any
gravitating media. The underlying tools in the study of the kinetics
of such media include, in particular, L\'{e}vy's steady
distributions (L\'{e}vy, 1937), whose special cases are the
power-law distributions like the Carpenter-de Vaucouleurs law in the
world of galaxies. Kinetic processes in such media are therefore
often referred to as L\'{e}vy processes, because they demonstrate
the tendency of the medium to approach some L\'{e}vy distribution.

Note, however,  that in contrast to  purely mathematical models,
where stochastic scale invariance can be infinite, all the real
fractal distributions, without exception show finite depth of the
hierarchy of scale invariance (scaling area). That is, these
distributions have the minimum and maximum scale, which limit the
applicability of fractal models. The key problem in the study of
particular real fractal distributions is the determination of the
limiting scale. This is not always an easy task, especially in
astrophysics, since these scales may lie outside the range of
available observational data. In the last two decades of the past
century and in the first decade of this century an especially
sharp and prolonged debate arose about determining the maximum
size for the fractal model of the distribution of galaxies in the
universe. The authors of different reputable studies found
estimates for this upper scale ranging from 5~to~600~$Mpc$. The
obvious reason for this variation was the lack of observational
data of sufficient accuracy and scope to solve this problem. Only
the latest data from the  WiggleZ Dark Energy Survey team
(Drinkwater et al., 2010) and the analysis of these data by
Scrimgeour et al. (2012) made it possible to close the issue.
Scrimgeour et al. (2012) used WiggleZ survey data in the
red-shift interval $0.1\leq z \leq 0.9$ to compute the correlation
dimension $D2$ (third dimension of Renyi fractal dimensions
spectrum). The above authors showed that the fractal dimension is
within $1\%$ of $D2 = 3$ for scale lengths exceeding $80h-1 Mpc$.
In other words, the distribution of matter in the universe can be
considered as practically uniform on the scale lengths greater
than $80h-1 Mpc$. Scrimgeour et al. (2012) also provide an
impressive historical review of the problem, to which we refer the
interested readers.

Chumak \& Rastorguev (2015) analysed Geneva-Copenhagen Survey of
local FG-type dwarfs (hereafter referred to as GCS) and showed
that the space distribution of stars of this type, which represent
a significant fraction of the local stellar population, differs
significantly from uniform Poisson distribution. We also showed
that the distribution of stars on GCS scales is consistent with
fractal model.

As is well known, the fundamental equation of stellar dynamics is
based on Poisson model (Binney \& Tremaine, 2008). Hence the
question arose whether it was the Poisson model is a proper tool
to analyze kinetic processes in stellar media. Certain progress
toward solving this problem was achieved by Vlad (1994), Chavanis
(2009), Chumak \& Rastorguev (2016), who generalized the Holtsmark
distribution to fractal media to derive the distribution of random
force acting on a test particle. The random force acting on a test
star was shown to be fully determined by the nearest neighbour
located at a certain {\it effective} distance $r_{m}$. Distance
$r_{m}$ is determined from the generalized nearest-neighbour
distribution. The average conditional density distribution was
derived and the fractal dimension of the near-solar stellar medium
was determined based on GCS data.

    In this paper we consider the motion of a test star in stellar
medium, and derive the fundamental equation of stellar dynamics that
takes into account, to a first approximation, the fundamental
properties of stellar gravitating media in terms of fractal model.

\section{EQUATION OF MOTION OF TEST STAR}
In the Lagrangian approach the equations of motion of test star $\it{i}$ in
the Galaxy in the local neighbourhood can be written in heliocentric
Cartesian Galactic coordinates in the following form:
\begin{equation}
\frac{dv_{ik}} {dt}=-av_{ik}+F^{rg}_{ik}+ F^{rg}_{ik}(t)
\end{equation}

where $k$ is used to indicate ${X, Y, Z}$ -- Cartesian coordinates
with the $X$-axis pointing toward the Galactic center; $a$ is the
coefficient of dynamic friction; $F^{rg}_{ik}$, the regular force,
and $F^{rg}_{ik}(t)$, the random (irregular) force produced by the
surrounding objects. For the sake of convenience we refer all
forces to a unit-mass test star.

In the case of local circumsolar volume considered here regular
force is time independent and reasonable estimates for the
components of this force can be derived in terms of the Oort
constants:
\begin{equation}
F^{rg}_{iX} \approx X_{i}A^{2},\quad F^{rg}_{iY} \approx
Y_{i}B^{2},\quad F^{rg}_{iZ} \approx 0.
\end{equation}

Equation set (1) can be rewritten in the form:
\begin{eqnarray}
\frac{dv_{ix}} {dt} & =& -av_{ix}+X_{i}A^2+ F^{ir}_{ix}(t)\nonumber \\
\frac{dv_{iy}} {dt} & =& -av_{iy}+Y_{i}B^2+ F^{ir}_{iy}(t)\\
\frac{dv_{iz}} {dt} & =& -av_{iz}+ F^{ir}_{iz}(t)\nonumber
\end{eqnarray}

Hereafter we consider the change of the test star velocity produced
by random force. In  fractal gravitating media the random force
acting on the test particle is practically fully determined by the
gravitational influence of the nearest neighbour whose {\it
effective} distance $r_{n}$ from the test star is given by the
following formula:
\begin{equation}
r_{n}=\frac{3}{D}(\frac{3}{4\pi h})^{1/D}\Gamma(\frac{D+1}{D})
\end{equation}

(Chumak \& Rastorguev, 2016). Here $D$ is the fractal dimension of
the medium; $\Gamma(x)$, the Gamma function, and $h$, the
normalization parameter of the mean conditional density $n(r)$ of
the medium:
\begin{equation}
n(r) = h r^{D-3}
\end{equation}

As is evident from equation (4), parameter $h$ has fractional
dimension equal to $[h]=L^{-D}$, where $L$  is the unit of length.
Hereafter throughout this paper we restrict ourselves, as we do in
equation~(5), to considering dimension $D$, i.e.,  the common
Hausdorff dimension, exclusively. This dimension is a rough
characteristic of the $n(r)$ distribution. The distribution $n(r)$
is actually multifractal, and $D$ corresponds to $D0$ of
multifractal spectrum of dimensions. This dimension describes the
structural heterogeneity of the distribution $n(r)$, which seems
to be the most significant in the framework of the present work.
For a more complete characterisation of the $n(r)$ distribution
one must, of course, calculate the entire spectrum of dimensions
$(D0, D1, D2,...)$ of the multifractal, or at least the first
three of these, which are the most important in practical
applications (see,for instance,(Chumak,2012)).

According to the above study, $r_{n}$ for the solar neighbourhood
is equal to a fraction of a parsec, and therefore when written in
the coordinate system connected with the center of mass of
interacting objects, equation set (3) contains small quantities on
the order of $r_n$ instead of $X_{i}$, $Y_{i}$, and $Z_{i}$, and
hence the effect of regular field on the act of encounter can be
neglected in the first approximation. The corresponding terms in
equation sets (1) and (3) can be dropped. Equation set (1) then
becomes:
\begin{equation}
\frac{dv_{ik}}{dt}=-av_{ik}+ F^{ir}_{k}(t).
\end{equation}

Hereafter $k=(x, y, z)$ are the coordinates of the star in the
reference frame connected with the center of mass. The medium is
locally isotropic on the scale length $r_{n}$ of the encounter and
therefore equations (6) are linearly independent. To simplify the
notation we drop the 'ir' superscript in the right-hand side and
the equation (6) takes the form:
\begin{equation}
\frac{dv_{ik}}{dt}=-av_{ik}+ F_{k}(t),
\end{equation}

The right-hand side of equation (7) consists of two terms: the force
of dynamical friction of the medium, which depends on the velocity
of the star and isotropic fluctuating force $F(t)$, which is
independent of the velocity of the test star. The formal solution of
equation (7) can be written as:
\begin{equation}
v_{ik}(t) = e^{-at}[v_{0ik} + \int^{t}_{0}e^{at'}F(t')dt'],
\end{equation}

where $v_{0ik}$  is the velocity of $i$-th star at time $t = 0$.

This solution is of no practical interest because the coefficient of
dynamic friction and random forces appearing to the right of the
integral sign are not written explicitly.

\section{DYNAMIC FRICTION}
Dynamic friction is determined by random two-point encounters, and
this fact allows the phenomenon to be analysed in terms of Markov
type processes. However, one has to take into account the dependence
of local density on scale length (5). As a result, the formula for
dynamic friction coefficient а in equation (7) can be rewritten as
follows, by dropping small terms on the order of $1/\lambda$ and
$p_{min}/p_{max}$ (see, e.g., Chapter 7 in the monograph of Binney
and Tremaine (2008)):
\begin{equation}
a \approx \frac{8\pi G^2 m^2 n \lambda} {v^3_{0i}}
\end{equation}

where $\lambda=ln(p_{min}/p_{max})$  is the Coulomb logarithm; $G$,
the gravitational constant; $m_{i}$, and $m$, the mass of the test
and field star, respectively; $n$, the number density of field
stars; $p_{min} = 2Gm/v^{2}_{i}$, the impact parameter of the close
encounter in which the test star is deflected by $\pi/2$ angle, and
$p_{max}$, the cutoff parameter for distant interactions (in our
case $p_{max}=2r_{m}$.

In this case the dynamical friction coefficient $a$ is computed by
analysing two-point encounters, because the average conditional
density $n$ (5) can be considered to be uniform and isotropic on
scale lengths on the order of $r_{n}$. We set $m_{i}=m$ to derive
the following formula for $a$:
\begin{equation}\label{eq:1}
a \approx \frac{4\pi G^2 m^2 h \lambda} {v^3_{0}}r^{(D-3)}.
\end{equation}

The dynamic friction coefficient $a$ of fractal and uniform media
differ fundamentally in that the former explicitly on scale length
$r$, because the corresponding formula contains conditional density
instead of uniform density. Correspondingly, there is no unique
dynamic friction coefficient for a fractal medium. The average
friction coefficient obviously depends on the volume considered, i.e., for
fractal media the definition of dynamic friction is also a matter of
convention. This is rather unusual; however, this is due to specific of
the kinetics in real fractal gravitating media: instead of fixed
average diffusion coefficients we have to deal with a continuous
spectrum of coefficient values. It is immediately apparent from
equation (10) that, as expected, it acquires the classical form in
the special case of uniform distribution:$D\rightarrow 3$;
$h\rightarrow n_{p}$.

Dynamic friction coefficient $а$ can approximately estimated using
the "correlation length" $r_{0}$ proposed by Davis \& Peebles (1983).
Thus
\begin{equation}\label{eq:2}
a_{0}=a(r_{0})\approx\frac{8\pi
G^{2}m^{2}h\lambda}{v_{0}^{2}}r_{0}^{(D-3)}\tag{\ref{eq:1}$'$}
\end{equation}

where $r_{с}$ can be determined from equality $n(r_{0}) = 1$ and, in
accordance with equation (5), it is equal to:

\begin{equation}\label{eq:3}
r_{0}=h^{-\frac{1}{D-3}} \tag{\ref{eq:1}$''$}
\end{equation}

\section{RANDOM FORCE}
According to Vlad (1994) and Chumak \& Rastorguev (2016), the
asymptotic distribution $W(|\vec{F}|;D)$  of the random force
$|\vec{F}|$ intensity for large forces in a medium with fractal
dimension can be written in the form:
\[
W(|\vec{F}|;D)d|\vec{F}| =
\]
\begin{equation}
=4\pi h(Gm)^{D/2}exp[-\frac{4\pi
h}{3}(\frac{GM}{|\vec{F}|})^{D/2}]|\vec{F}|^{-\frac{D+2}{2}d|\vec{F}|}
\end{equation}

Let us rewrite formula (11) using a more compact notation:
\begin{equation}
W(F_{m};D)dF_{m} =3 b \cdot exp(-b\cdot F_{m}^{-D/2})dF_{m},
\end{equation}

where, to simplify the formula, we adopted $|\vec{F}|=F_{m}$,
\begin{equation}
b=\frac{4\pi h}{3}(Gm)^{D/2}.
\end{equation}

Distribution (11) is stationary. Let us assume that there are no
temporal correlations in the random process of the motion of a
test star and all directions of random force are equally
represented. Subscript 'k' can then be dropped in the force terms
in equations (7) and (8) and $F_k (t) = F (t)$ can be written as
the product of:
\begin{equation}
F(t;t_c)=F_{m}\varphi_{i}(t;t_{c}),
\end{equation}

where $t_{c}$ is the time scale over which the test star interacts
(collides) with a field object. Let us assume that the collision
(closest encounter) of $i$-th star occurs at time $t'$. At that time
the force is maximal. Suppose, for simplicity, that the force acting
on the star is symmetric with respect to time $t= t'$  and decreases
with time in  accordance with the following law:
\begin{equation}
F(t;t_c)=\frac{F_{m}}{t_{c}\sqrt{\pi}}e^{-(\frac{t-t_{c}}{t_{c}})^{2}}.
\end{equation}

If $t_{c} << l/|v_{i}|$, where $l$ is the characteristic free path,
we let $t_{c}$ to zero to derive the following formula instead of
equation (15):
\begin{equation}
F(t)=F_{m}\delta(t-t').
\end{equation}

Function $\varphi_{i}(t)$ then describes a delta-correlated random
process and, correspondingly, random forces $F(t)$ in equation (14)
satisfy the following conditions:
\[
\langle F(t)\rangle=0,
\]
\begin{equation}
\langle F(t)F(t')\rangle\approx[W(F_{m};D)F_{m}]^{2}\delta(t-t'),
\end{equation}

where angular braces denote averaging over the ensemble consisting
of $n$ independent realizations. For example,
\begin{equation}
\langle\varphi(x)\rangle=\frac{1}{n}\sum_{i=1}^{n}\varphi_{i}(x).
\end{equation}

Each realization represents a random element of stellar medium
containing one test star.

Formulae (17) have simple physical meaning: interaction is
practically instantaneous and all interactions are mutually
uncorrelated. Equation (7) in this case is the classical Langevin
equation. Below we apply the appropriate procedures and methods
described, e.g., in monographs of R\'{e}sibois and Leener (1982) and
van Kampen (1990) and other studies, to solve this equation and
perform its subsequent analysis.

\section{DIFFUSION COEFFICIENTS}
Given the first condition (17), we conclude from formal solution (7)
that
\begin{equation}
\langle v_k(t)\rangle = v_{0k}e^{-at},
\end{equation}

where $v_{0k}$  is the initial velocity component of the test star
at initial time $t = 0$.

We then multiply equation (8) by a similar expression, and, given
that $\langle v_{k}(t)\cdot v_{l}(t)\rangle=0$ for $k\neq l$, and
$\langle v_{k}(t)\cdot v_{k}(t)\rangle = \langle
v^{2}_{k}(t)\rangle$, we obtain:
\begin{equation}
\langle v^{2}_{k}(t) \rangle=v_{0k}^{2}e^{-2at}
\int_{0}^{t}dt'\int_{0}^{t'}dt'' e^{+a(t'+t'')}\langle
F(t')F(t'')\rangle.
\end{equation}

Hereafter subscript 'i' is not due to averaging over the ensemble
of realizations (see. (18) as an example). Given the second
condition (17) we obtain, after performing the necessary
operations:
\begin{equation}\label{eq:4}
\langle v^{2}_{k}(t)\rangle =
v_{0k}^{2}e^{-2at}+\frac{1}{2a^{2}}[W(F_{m};D)F_{m}]^{2}(1-e^{-2at}).
\end{equation}

The expression$W(F_{v};D)F_{m}dF_{m}$ is equal to the probability
of finding in the ensemble of two-point interactions considered
here, a random field intensity in the interval $F_{m}+dF_{m}$
subject to the asymptotic form of generalized Holtsmark
distribution (12). For numerical estimates the expression in
squared brackets should be replaced by some value averaged over
the ensemble of physically realizable $F_{m}$:
\begin{equation}
\langle F_{m}\rangle =
\int_{F_{min}}^{F_{max}}W(F_{m};D)F_{m}dF_{m},
\end{equation}

where $F_{max}$ and $F_{min}$ are the boundaries of the interval of
such physically realizable $F_{m}$ values. We now rewrite formula
(11) in terms of dimensionless variable $x$:
\begin{equation}
x = \frac{4\pi h}{3}(\frac{Gm}{|\vec{F}|}) = \frac{4\pi
h}{3}(\frac{Gm}{F_{m}})^{D/2}.
\end{equation}

Formula (22) then becomes
\begin{equation}
\langle F_{m} \rangle = \gamma_{D}F_{0}, \; \;  \gamma_{D}=\int_{x_{min}}^{x_{max}}e^{-x}x^{-2/D}dx,
\end{equation}

where
\[
F_{0}=
\frac{2Gm}{D}(\frac{4\pi h}{3})^{2/D}
\]
and
\begin{equation}
x_{min}=\frac{4\pi h}{3}p_{min}^{D},  x_{max}= \frac{4\pi
h}{3}p_{max}^{D}.
\end{equation}

Here, like in formula (9),  $p_{min}= 2Gm / v_{i}^{2}$,
  $p_{max}\approx 2r_{n}$.
The integral in formula (24) for $x \rightarrow \infty$  can be
written in terms of Whittaker function $W_{\delta,\lambda}(u)$:
\begin{equation}
\langle F_{m} \rangle =
F_{0}x^{-\frac{1}{D}}e^{-\frac{x_{min}}{2}}W_{-\frac{1}{D},\frac{(D-1)}{2D}}(x_{min})
\end{equation}

In the particular cases $D = 1$ and $D = 2$, integral (24) can be
expressed in terms of hypergeometric function $Ei(x)$. For arbitrary
$x_{min}$, $x_{max}$, and $D$ integral (24) has to be computed
numerically.
    We finally rewrite formula (21) in the form:
\begin{equation}\label{eq:5}
\langle v^{2}_{k}(t) \rangle =
v_{0}^{2}e^{-2at}+\frac{(1-e^{-2at})}{2a^{2}}\langle
F_{m}\rangle^{2} \tag{\ref{eq:4}$'$}
\end{equation}

We then use equation (19)  to compute the time scale $\tau _{fr}$ of
the deceleration of a star due to dynamic friction:
\begin{equation}\label{eq:6}
\frac{d\langle v(t) \rangle}{dt}\sim\frac{\langle v(t) \rangle}{\tau
_{fr}}\rightarrow \tau _{fr}=\frac{\langle v(t) \rangle}{d\langle
v(t) \rangle /dt}=a^{-1}.
\end{equation}

It is evident from formula (21') that
\begin{equation}
\sigma_{ec}^{2}= \lim_{t\rightarrow \infty}\langle v^{2}(t)
\rangle\rightarrow\frac{\langle F_{m} \rangle^{2}}{2a^{2}}.
\end{equation}

Physically this means that whereas in this limit dynamic friction
(19) reduces the initial velocity of test stars to zero, diffusion
due to random force occurs independently of initial velocity and
leads to the equilibrium dispersion $\sigma^{2}_{ес}$ determined
by formula (29). As a consequence of these two processes the test
ensemble of stars completely "forgets" its initial conditions, and
the final dispersion of the ensemble becomes equal to the
equilibrium velocity dispersion as defined by equation (28). Let
us denote the distribution function of the equilibrium state of
fractal star environment as $f_0$. Consider a small deviation of
the distribution function $f$ from the equilibrium state $f_0$. In
this case, by analogy with (27) and taking into account equation
(28), we obtain the following formula for the characteristic
diffusion time $\tau_df$:
\begin{equation} \label{eq:7}
\tau_{df}=\frac{\sigma_{ec}^{2}-\langle v^2_k (t)\rangle}{d\langle
v^2_k (t)\rangle / dt}=(2a)^{-1} \tag{\ref{eq:6} $'$}
\end{equation}

In view of the fluctuation-dissipation theorem in its simplest form,
we conclude that $\sigma^{2}_{ес} = \sigma_{0}^{2})$  is the average
local velocity dispersion determined in the observable volume of
size $r_{ob}$, whereas parameter $a$ is, according to equation (10),
a function of $r$. In view of formulae (10) and (24), after simple
manipulations we derive from equation (27') the following formula:
\begin{equation}
\sigma_{0}^{2}=2(\frac{4\pi}{3})^{\frac{4}{D}}(\frac{Gm\gamma_{D}}{D\sqrt{a_{1}}})^{2}h^{\frac{4-D}{D}}r^{3-D}
\end{equation}

where
\[
a_{1}=\frac{8\pi G^{2}m^{2}\lambda}{v_{0}^{3}},
\]
and the value of integral $\gamma_D$ and other constants were
determined above.

It follows from formula (29) that the cause of the existence of
equilibrium velocity dispersion in a stellar medium is the
non-Poissonian (nonuniform) structure of this medium. We pass to
Poisson limit in equation (29) $(D\rightarrow 3)$  to derive a
formula for equilibrium velocity dispersion in a uniform Poisson
medium:
\begin{equation}
\sigma_{0,D\rightarrow
3}^{2}=2(\frac{4\pi}{3})^{\frac{4}{3}}(\frac{Gm\gamma_{3}}{3\sqrt{a_{1}}})^{2}\sqrt[3]n
\end{equation}

We then determine $\gamma_{0}$ from equation (24) by passing
everywhere to Poisson limit.

Accurate numerical estimates by formula (29) make it possible, in
principle, to obtain an independent estimate of the fractal
dimension $D$ of the medium considered, and a comparison of the
dispersion estimates given by formulae (29) and (30) can be used to
estimate the contribution of fractal inhomogeneities to the total
dispersion.

\section{FUNDAMENTAL EQUATION OF STELLAR DYNAMICS IN THE FOKKER-PLANCK APPROXIMATION}
Fundamental equation of stellar dynamics - Boltzmann's kinetic
equation with collisional term - can be written in the following
form (see, e.g., formula (5.3) in Ogorodnikov (1958)):
\begin{equation}
\frac{\partial f}{\partial t}+v_{k}\frac{\partial f}{\partial
x_{k}}+\dot{v}_{k}\frac{\partial f}{\partial v_{k}}=(\frac{\partial
f}{\partial t})_{ir}.
\end{equation}

Here $f (t, x_{k}, v_{k})$ is the distribution function of stars
of the test ensemble like in equation (1) above; $k$ denotes ${X,
Y, Z}$; summation is performed over repeated indices; dot denotes
differentiation with respect to time, and the term in the
right-hand side of the equation describes the change of the
distribution function due to irregular forces (the collisional
term). The second and third terms in the left-hand side of
equation (31) are determined by the model of the stellar system
and their characteristic spatial and temporal scales are
substantially greater than the distances between stars and typical
time scales of random-force variations, respectively, that appear
in computations of the right-hand side of equation (31). We, like
usually, assume that the second and third terms of the left-hand
side of equation (31) are known, and focus on the change of the
distribution function due to random forces. Let us consider this
effect in the Fokker-Planck approximation.

As is well known, the Fokker-Planck approach describes in the first
approximation the process of diffusion in the velocity space due to
the influence of fluctuating random force on the ensemble of test
particles, causing small successive changes of particle velocities.
In this approximation the time derivative of phase-space density or
the collision term $(\partial f/\partial t)_{ir}$ that appears in
the right-hand side of equation (31), is viewed as an additive part
of the total divergence of flow $\vec{J}_{ir}$ in the velocity space
due to random force:
\begin{equation}
(\frac{\partial f}{\partial t})_{ir} =
-div_{v}\vec{J}_{ir}=-\frac{\partial}{\partial
v_{k}}({\dot{v}_{k}f}),
\end{equation}

where $\dot{v}_{k}$  is the acceleration produced by random force.

In the Fokker-Planck approximation the usual practise is to expand
flow $\vec{J}{J}_{ir}$ into a series and leave only its first two
terms:
\begin{equation}
J^{^{ir}}_{k}=c_{k}f+c_{kl}\frac{\partial f}{\partial
v_{k}}+c_{klm}\frac{\partial^{2}f}{\partial v_{k}\partial v_{l}}+...
\approx \dot{v}_{k}f+\dot{v}_{kl}\frac{\partial f}{\partial v_{k}}.
\end{equation}

The accelerations due to random forces and appearing in the
right-hand side of equation (32) can be determined from formulas
(19) and (21'). Let is consider the variations of our solutions (19)
and (21') with time over small time scales on the order of $\Delta t
<< a^{-1}$. We restrict the expansion of exponential function in
formula (19) to the first two terms to obtain:
\begin{equation}
\dot{v}\approx \frac{\Delta v}{\Delta t}=\frac{\langle v(t) \rangle
- v_{0}}{\Delta t} = - a v_{0}+\varepsilon (\Delta t)^{2}.
\end{equation}

We similarly derive from equation (21'):
\begin{equation}
\dot{v}_{kl}= \frac{\Delta v^{2}_{k}}{\Delta t}=\frac{\langle
v^{2}_{k}(t) \rangle - v_{0k}^{2}}{\Delta t} =
-2av_{0k}^{2}+a^{-1}\langle F_{m} \rangle^{2}+\varepsilon (\Delta
t)^{2}.
\end{equation}

Here the problem reduces to one dimension and therefore the
symmetric second-rank tensor $\dot{v}_{ij}$  is represented by its
single trace term.

We now drop the second-order terms in $\Delta t$ in formulae (35)
and (35) and take into account formulae (33) and (34) to write the
term in the right-hand side of equation (32) in the following form:

\begin{eqnarray}
(\frac{\partial f}{\partial t})_{ir}=av_{0k}\frac{\partial
f}{\partial v_{k}}+(2av_{0k}^{2}-a^{-1}\langle F_{m}
\rangle^{2})\frac{\partial^{2}f}{\partial v^{2}_{k}} = \nonumber \\
= av_{k}\frac{\partial f}{\partial v_{k}}+2a(v_{k}^{2}-
\sigma_{eq}^{2})\frac{\partial^{2}f}{\partial v^{2}_{k}}
\end{eqnarray}

Here we omit the '0' subscript at the test-star velocity in the
right-hand side, because an arbitrary field star can be used as a
test particle with no loss of generality. The collisional term in
the Fokker-Planck approximation indicates the direction and time
scales of kinetic processes at different stages of the evolution
of the test ensemble. In particular, in the limit (21$'$), where
the system "forgets" its initial conditions, formula (36)
simplifies and becomes

\begin{equation}
\lim_{t\rightarrow\infty}(\frac{\partial f}{\partial
t})_{ir}=a^{-1}\langle F_{m} \rangle^{2}\frac{\partial
^{2}f}{\partial v^{2}_{k}}=2a\sigma^{2}_{eq}\frac{\partial
^{2}f}{\partial v^{2}_{k}}
\end{equation}

Equation (31) with the right-hand side of the form (36) can be
viewed as the fundamental equation of stellar  dynamics in the
Fokker--Plank approximation. Its last form in the rightmost-hand
side in formula (37) can be used in the case of very small
deviations of fractal gravitating medium from equilibrium. One of
the main difficulties in solving kinetic equation (31) is due to
the complex structure of collision integral (36). Approximation
(37) also does not eliminate these difficulties. It is a common
practise when dealing with equations of the form (31) to use the
so-called model collision integrals - simpler operators
preserving, without the fine details, the basic meaning and the
main qualitative properties of the original exact collision
integral. Equations of the form (31) with model collision
integrals  are usually referred to as model kinetic equations. One
of the most famous equations of this kind is the BGK equation
(Bhatnagar et al., 1954). If the deviation from the equilibrium
distribution $f_{0}$ is small, then in the approximation of BGK
model equation we have:

\begin{equation}
(\frac{\partial f}{\partial t})_{ir} \approx \frac{f-f_{0}}{\tau}
\end{equation}

where, given equations (27$'$), (28), we have
\begin{equation}
\tau =\tau_{df}=\frac{1}{2a}=\frac{a \sigma_{es}^{2}}{\langle F_{m}
\rangle}^{2}.
\end{equation}

The approach represented by formulae (34) and (35) yields for
$\tau_{}fr$ the estimate identical to that given by formula (27).
For $\tau_{fr}$ we get another expression:

\begin{eqnarray}
\dot{v_{kl}}=\frac{\langle v_k^2 \rangle -v_{0k}^2}{\tau_{df}} =
-2av_{0k}^2 + a^{-1}\langle F_m^2 \rangle \rightarrow \nonumber \\
\tau_{df} = \frac{\langle v_k^2 \rangle
-v_{0k}^2}{2a(\sigma_{eq}^2 - v_{0k}^2)}
\end{eqnarray}

It is evident from equation (40) that in the case $v_{0k}^2
\rightarrow \sigma_{eq}^2$ we have $\tau_{df}, \; v_{0k}^2
\rightarrow \infty $. In the opposite case, $v_{0k}^2 \rightarrow
0$, we obtain, in view of equations (21) and (40),

\begin{equation*}
\tau_{df}, \; v_{0k}^2 \approx \frac{1}{2}(1 - e^{-2at})
\rightarrow 0
\end{equation*}
for small $t$. We pointed out a similar behaviour of
characteristic time scales and velocity dispersion in our earlier
paper (Chumak \& Rastorguev 2014), where we derived a direct
solution of Landau equation for non-equilibrium stellar media. On
the other hand, if $\langle v_{0k}^2 (t) \rangle \rightarrow
\sigma_{eq}^2$, we obtain the same result as presented in equation
(27) and, respectively, obtain the time-scale ratio
$\tau_{df}/\tau_{fr}$=2. Since we use approximation (36) and BGK
approximation (38), it is preferable to use the $\tau_{df}$ value
from formula (39).

Strictly speaking, the BGK approximation was proposed as a tool to
derive qualitative estimates for solving the Boltzmann equation.
However, it appears that the application of model collision
integral (38) is also justified in this case, because here we
consider the transition of the medium to stable equilibrium
distribution, which is the main condition for the application of
the model BGK approximation (Cercignani, 1975). Equation (31) with
approximate collision integral  (38) allows one to derive integral
equations for the hydrodynamic variables in any problem
considered. Many interesting problems require solving these
equation. Note, however, that this approach can give only a
qualitative estimate. Whenever possible, such solutions should be
verified by direct numerical simulation, or by solving equation
(31) with collision integral (36).

\section{DISCUSSION}
To demonstrate significant differences of the fractal model of
gravitating medium from the Poisson model generally adopted in
these models numerically estimated for the local solar
neighbourhood based on the results of a fractal analysis of the
GCS. We adopt $D \approx 1.23$ and $h\approx 1.644$ for the
distribution of conditional density in the fractal medium from our
earlier paper (Chumak \& Rastorguev, 2015).

Formula (3) allows us to estimate the average effective distance
to the nearest neighbour. We obtain $r_{f}\approx 0.48$ pc for the
fractal model. The corresponding estimate for the Poisson model
with $D \approx 3$ and $h \approx n \approx 0.10$ $pc^{-3}$ (we
adopt local stellar density from Binney \& Tremaine, 2008) is
$r_{р} \approx 1$ pc, i.e., almost twice greater than for the
fractal case.

Formula $r_{0}=h^{-\frac{1}{D-3}}$ yields the "correlation length"
of $r_{o} \approx 2.41$ pc for the fractal model, which is about
five times the effective interparticle distance. In the Poisson
model such a parameter does not exist by definition.

We derive from equation (10) the following formula for the time
scale of test star deceleration due to dynamic friction of fractal
medium:
\begin{equation}
\tau_{f}=\frac{v_{0}^{3}}{8\pi G^{2}m^{2}h \lambda}r^{(D-3)}.
\end{equation}

In the Poisson limit $(D\rightarrow 3; h \rightarrow n_{p})$ we
obtain the well-known formula for the relaxation time
\begin{equation}
\tau_{p}=\frac{v_{0}^{3}}{8\pi G^{2}m^{2}n \lambda}.
\end{equation}

Hence
\begin{equation}
\tau_{fr}=n h^{-1}r^{(D-3)}\tau_{p}.
\end{equation}

We substitute into this formula the parameter values for the solar
neighbourhood mentioned above and adopt $r = r_{0}$ to obtain
$\tau_{f} \approx 0.032\tau_{p}$. In other words, on this scale
the characteristic time of dynamic deceleration in terms of
fractal model is about a factor of 30 shorter than in the Poisson
limit. As is evident from formula (42), this characteristic time
is even shorter on scale lengths $r < r_{0}$. It becomes equal to
the Poisson-limit time scale only on scale lengths of $\sim 300$
pc. At greater distances $\tau_{fr} > \tau_{p}$.

In this framework of this approach there exist small deviation
from the current dispersion from its equilibrium value, $\langle
v_{0k}^2 (t) \rangle \rightarrow \sigma_{eq}^2$, and we obtain the
relation $\tau_{df} = 2 \tau{fr}$. However, it should be noted
that for larger deviations from the equilibrium formula (39), the
ratio between the characteristic times will differ from that, and,
accordingly, an approximate model (38) is not applicable. In these
cases, the approximation (34), (35) is no more applicable, and we
should use more realistic approximation of the collision integral.

And one more point. The kinetics of gravitating media differs
fundamentally from the that of uniform media (on scale lengths
that are long compared to interparticle distance) because in the
former case the diffusion coefficients appearing in kinetic
equations depend on spatial scale length. This difference is
emphasized quite clearly in the approach to kinetics developed by
a number of authors and in this paper, involving averaged
conditional density and aver- age fractal dimension on the scale
length of the fragment of the gravitating medium studied. However,
this is by no means the only difference. Be- cause of its
alternating pattern the structure of fractal medium with all its
multiscale concentrations and voids is much more complex. To
establish a consistent kinetic theory of such a medium, a special
mathematical apparatus has to be developed that should free of
such strong assumptions as the validity of using average
conditional density and aver- age dimension. The deviations of the
actual values of these quantities from their means may be quite
significant. Therefore our results should be viewed only as a
first approximation, which appears to be closer to reality than
the classical approximation based on the model of uniform Poisson
space distribution. An alternative approach to the development of
kinetics of fractal media is based on the derivation of fractional
kinetic equations for each particular problem (see, e.g., Saichev
\& Zaslavsky, 1997). See also the monograph by Uchaikin (2008) for
a review of publications in this field, which includes more than
one hundred references. Fractional generalization of kinetic
equations is based on the use of fractional derivatives with
respect to time and coordinates. This technique will possibly
produce more accurate approximations for the problem considered in
this paper.

\section*{ACKNOWLEDGMENTS}
A.S.~Rastorguev acknowledges partial support from the Russian
Science Foundation (grant no. 14-22-00041) and O.V.~Chumak
acknowledges the support from the Russian Foundation for Basic
Research (grant no. № 14-02-00471). The authors are also grateful
to anonymous referee for valuable notes and advice.


\begin{thebibliography}{99}

\bibitem[\protect\citeauthoryear{Benhaiem et al.}{2013}]{b35} Benhaiem D., Joyce M., Sicard F., 2013, MNRAS,
429, 3423

\bibitem[\protect\citeauthoryear{Binney\&Tremaine}{2008}]{b1} Binney J., Tremaine S., 2008, Galactic Dynamics,
Princeton University Press

\bibitem[\protect\citeauthoryear{Bhatnagaret al.}{1954}]{b2} Bhatnagar P.L., Gross E.P., Krook M., 1954, Phys. Rev., 94, 511

\bibitem[\protect\citeauthoryear{Carpenter}{1938}]{b3} Carpenter E.F., 1938, Astrophys. J., 88, 344

\bibitem[\protect\citeauthoryear{Chavanis}{2009}]{b4} Chavanis P.H., 2009, Eur. Phys. J. B, 70, 413

\bibitem[\protect\citeauthoryear{Cercignani}{1975}]{b5} Cercignani C., 1975, Theory and Application of the Boltzmann
Equations , Scottish Acad. Press. Edinburg and London.

\bibitem[\protect\citeauthoryear {Chumak\&Rastorguev}{2015}]{b6} Chumak O.V., Rastorguev A.S., 2015, Baltic Astron., 24, 1, 302

\bibitem[\protect\citeauthoryear {Chumak\&Rastorguev}{2016}]{b7} Chumak O.V., Rastrorguev A.S., 2016, Astron. Letters, 42, 5, 307

\bibitem[\protect\citeauthoryear {Chumak}{2012}]{b8} Chumak O. V., 2012, Entropies and fractals in data analysis,
R\&C Dynamics. Moscow-Izhevsk, p.166 (in Russian)

\bibitem[\protect\citeauthoryear {Davis\&Peebles}{1983}]{b9} Davis M., Peebles P.J.E., 1983, Astrophys. J., 267, 465

\bibitem[\protect\citeauthoryear {Drinkwater et al.}{2010}]{b10} Drinkwater M. J. et al., 2010, MNRAS, 401, 1429

\bibitem[\protect\citeauthoryear {Einasto}{1989}]{b11} Einasto J., 1989, in Proc. 3rd ESO-CERN Symp., Astronomy, Cosmology
and Fundamental Physics (ed. M. Cato), Dordrecht: Kluwer, 231

\bibitem[\protect\citeauthoryear {Joyce et al.}{1999}]{b12} Joyce M., Montuori M., Labini F.S., 1999, Astrophys. J., 514, L5

\bibitem[\protect\citeauthoryear {Joyce \& Sicard}{2011}]{b34} Joyce M., Sicard F., 2011, MNRAS, 413,
1439

\bibitem[\protect\citeauthoryear {Klypin et al.}{1989}]{b13}Klypin A. A., Einasto J., Einasto M., Saar E., 1989, MNRAS, 237, 929

\bibitem[\protect\citeauthoryear {Labini et al.}{1996}]{b14} Labini F.S. Gabrielli A., Montuori M., Pietronero L., 1996, Physica
A: Statistical Mechanics and its Applications, 226, 195

\bibitem[\protect\citeauthoryear {Larson}{1981}]{b15} Larson R.B.,  1981, MNRAS, 194, 809

\bibitem[\protect\citeauthoryear {L\'{e}vy}{1937}]{b16} L\'{e}vy P., 1937, Th\'{e}orie de l'Addition des Variables
Al\'{e}atoires, Paris: Gauthier-Villars

\bibitem[\protect\citeauthoryear {Mandelbrot}{1977}]{b17} Mandelbrot, B.B., 1977, Fractals: Form, Chance and Dimension, W.H.
Freedman

\bibitem[\protect\citeauthoryear {Mandelbrot}{1987}]{b18} Mandelbrot B.B., 1987, Self-affine fractal sets in Fractals in
Physics, Amsterdam: North-Holland Physics Publishing, 3

\bibitem[\protect\citeauthoryear {Miller \& Rouet}{2010}]{b33} Miller B.N., Rouet J.-L., 2010, Journal of Statistical
Mechanics: Theory and Experiment, Issue 12, 12028

\bibitem[\protect\citeauthoryear {Ogorodnikov}{1958}]{b19} Ogorodnikov K.F., 1958, Dynamics of stellar systems, Moscow: Glavnoe
Izd. Fiz. Mat. Lit.

\bibitem[\protect\citeauthoryear {Peebles}{1980}]{b20} Peebles P. J. E., 1980, The Large-Scale Structure of the Universe.
Princeton Univ. Press

\bibitem[\protect\citeauthoryear {Peebles}{1993}]{b21} Peebles P. J. E., 1993, Principles of Physical Cosmology. Princeton
Univ. Press

\bibitem[\protect\citeauthoryear {Pietronero et al.}{2002}]{b22} Pietronero L., Gabrielli A., Labini F.S., 2002, Physica A:
Statistical Mechanics and its Applications, 306, 395

\bibitem[\protect\citeauthoryear {Pietronero \& Labini}{2005}]{b23} Pietronero L., Labini F.S., 2005, Complexity, metastability and nonextensivity (31st Workshop of the International School of Solid
State Physics, held 20-26 July 2004 in Erice, Sicily, Italy; Eds.
C. Beck, G. Benedek, A. Rapisarda, C. Tsallis), World Scientific
Publishing Co. Pte. Ltd., 91

\bibitem[\protect\citeauthoryear {R\'{e}sibois\&Leener}{1977}]{b24} R\'{e}sibois P., de Leener M., 1977, Classical Kinetic Theory of
Liquids. N.Y., Wiley

\bibitem[\protect\citeauthoryear {Rouet et al.}{1990}]{b37} Rouet J.-L., Feix M.R., Navet
M., 1990, Vistas in Astronomy, 33, 357

\bibitem[\protect\citeauthoryear {Saichev \& Zaslavsky}{1997}]{b25} Saichev A. I.,Zaslavsky G. M., 1997, Chaos, 7, 753

\bibitem[\protect\citeauthoryear {Scrimgeour}{2012}]{b26} Scrimgeour M.I., et al., 2012, MNRAS, 425, 116

\bibitem[\protect\citeauthoryear {Shapley}{1934}]{b27} Shapley H., 1934, MNRAS, 94, 791

\bibitem[\protect\citeauthoryear {Shiozawa \& Miller}{2016}]{b36} Shiozawa Y., Miller B.N., 2016, Chaos, Solitons and
Fractals, 91, 86

\bibitem[\protect\citeauthoryear{van Kampen}{2007}]{b28} van Kampen N.G., 2007,
Stochastic Processes in Physics and Chemistry. North Holland

\bibitem[\protect\citeauthoryear {Uchaikin}{2008}]{b29} Uchaikin V.V., 2008, Method of fractional derivatives. Ulyanovsk:
"Artishok", (in Russian)

\bibitem[\protect\citeauthoryearde {Vaucouleurs}{1970}]{b30} Vaucouleurs, G., 1970, Science, 167, Is. 3922, 1203

\bibitem[\protect\citeauthoryearde {Vega et al.}{1998}]{b31} Vega H. J., Sanchez N., Combes  F., 1998, Astrophys. J., 500, 8

\bibitem[\protect\citeauthoryear {Vlad}{1994}]{b32} Vlad M.O., 1994, Astrophys. and Space Sci., 218, 159

\end{thebibliography}
\end{document}